\documentclass[aps,pre,superscriptaddress,floatfix]{revtex4-2} 
\usepackage{graphicx}   
\usepackage{amsmath}    
\usepackage{amssymb}    
\usepackage{hyperref}   
\usepackage{bm}         
\usepackage{subfigure}
\usepackage{soul}       
\usepackage{xcolor}     
\sethlcolor{yellow} 
\begin{document}

\title{Parametric Modulation of Nonlinear Coupling in the Hénon–Heiles System: Resonances, Chaos, and Stabilization}

\author{Vinesh Vijayan}%
\affiliation{Department of Science \& Humanities, Rathinam Technical Campus, Coimbatore, India}
\email{vinesh.physics@rathinam.in}
\author{P Satishkumar}%
\affiliation{Department of Mechanical Engineering, Rathinam Technical Campus, Coimbatore, India}
\author{Dinesh Wankhade}%
\affiliation{Department of Electrical Engineering, St: Vincent Pallotti College of Engineering, Nagpur, India}
\author{R Sumathi}
\affiliation{Department of Science \& Humanities, Rathinam Technical Campus, Coimbatore, India} 
\date{\today}

\begin{abstract}
We investigate parametric modulation of the nonlinear coupling in the Hénon--Heiles system, which directly modifies intrinsic resonance structure in a manner complementary to additive forcing. Canonical perturbation theory in extended phase space yields normal forms predicting resonance tongues scaling as $\sqrt{\varepsilon}$ near commensurate frequencies. Melnikov analysis quantifies separatrix splitting and chaos onset, confirmed by symplectic simulations showing transition from localized resonances to global transport via overlap. High-frequency averaging reveals potential stiffening that suppresses chaos. parametric modulation of nonlinear coupling provides an alternative route for generating combination resonances and influencing chaotic dynamics
\end{abstract}

\maketitle
\section{Introduction}
The Hénon–Heiles system, a benchmark model in the study of the transition to chaos in Hamiltonian dynamics, exhibits a dramatic change in behavior characteristic of nonlinear systems. Originally introduced to describe the motion of a star near the galactic center, the model reveals a transition from regular to chaotic motion at a critical energy$(\bf{E_c=\frac{1}{6}})$, determined by the structure of the potential energy surface and the topology of the corresponding equipotential contours. This transition is a hallmark of the system's rich dynamical behavior. The Hamiltonian for this two-dimensional model is given by:
\begin{equation}
H=\frac{1}{2}(p_x^2+p_y^2)+\frac{1}{2}(x^2+y^2)+\lambda(x^2y-\frac{1}{3}y^3)
\label{E1}
\end{equation}
Here, \( x \) and \( y \) denote the spatial coordinates, and \( p_x \) and \( p_y \) are their respective conjugate momenta. The parameter \( \lambda \), typically set to unity, governs the strength of the nonlinear terms. The \( x^2 y \) term introduces coupling between the \( x \) and \( y \) degrees of freedom, while the \( y^3 \) term contributes a cubic nonlinearity to the potential. Notably, the system is invariant under rotations by \( 120^\circ \) in the \( x\text{--}y \) plane, reflecting its underlying discrete rotational symmetry\cite{Mic}\cite{Con}\cite{Lic}. Parametric excitation within Hamiltonian systems has been widely explored in the realm of periodically driven oscillators, where time-varying parameters lead to instability regions and Floquet-like behavior~\cite{landau_mechanics_1976,nayfeh_mook_review_1979}. Such phenomena are generally examined in linear or mildly nonlinear setups, in which the modulation modifies the effective rigidity or oscillation frequency.
\\
\\
Subsequent studies on the Hénon–Heiles system have revealed a wealth of insights into its intricate dynamical behavior. Notably, the fractal structure of the exit basins has been associated with anomalous transport phenomena~\cite{Zas}. High-precision investigations employing the Smaller Alignment Index (SALI) have been instrumental in distinguishing between regular and chaotic motion~\cite{Bar}. The topology of chaotic trajectories and the occurrence of bifurcations, as elucidated through Poincaré surface-of-section maps, has also been examined in detail~\cite{Sai}. The open Hénon–Heiles system has been analyzed in connection with its relevance to other models, broadening its applicability~\cite{Tel}. A seminal contribution offers a comprehensive treatment of the invariant manifolds, providing a rigorous mathematical framework~\cite{Wig}. Moreover, the study of escape dynamics and the unpredictability of exit basins continues to be a compelling area of research~\cite{deM}. Finally, the system serves as a benchmark for testing symplectic integration schemes, particularly in the context of long-term energy conservation in numerical simulations~\cite{Yos}.\\
\begin{figure}[t]
    \centering
    \includegraphics[width=\textwidth]{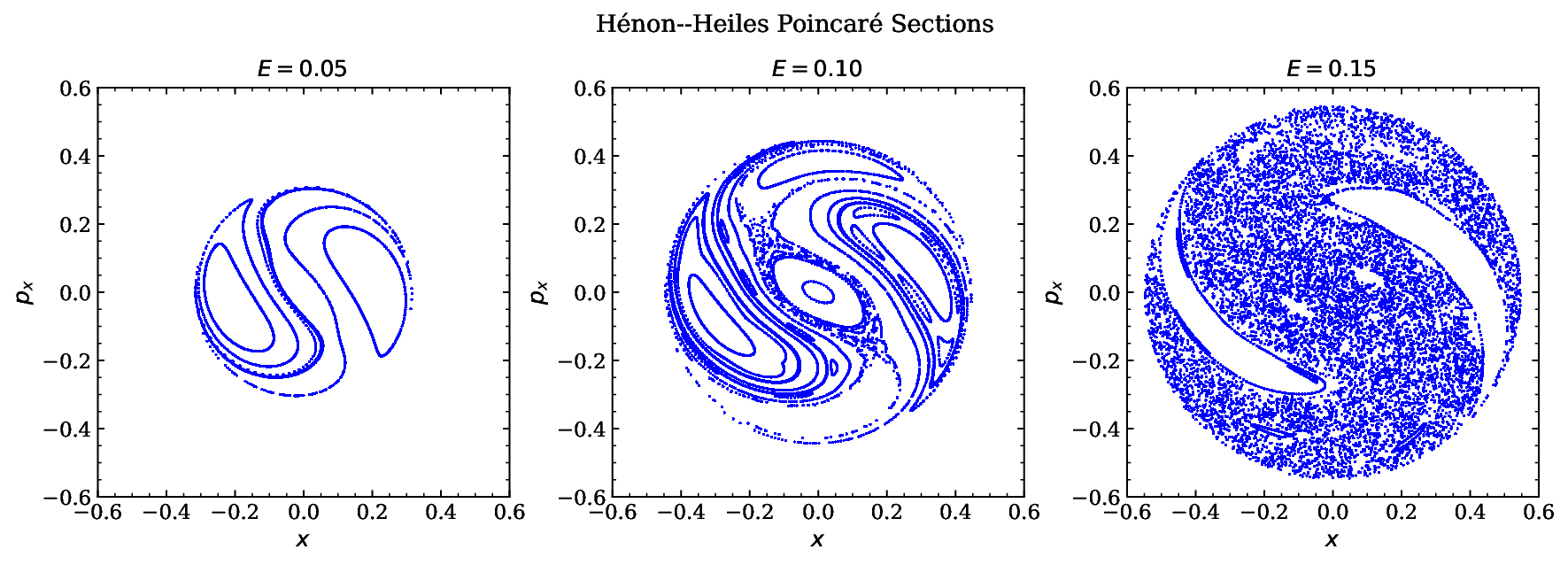}
    \caption{
        \textbf{Baseline Hénon--Heiles Poincaré section.} Poincaré map (\(y = 0,\, p_y > 0\)) for the unmodulated Hénon--Heiles system \(\lambda = 1\)), showing three energies below the escape threshold         displayed side by side. Each panel corresponds to one of the energies \(E = 0.05,\ 0.10,\ 0.15\), respectively.  
Trajectories were integrated using a symplectic velocity--Verlet method, and intersections with the Poincaré surface were recorded using linear interpolation. These baseline sections illustrate the regular island structure and the growth of chaotic regions as the energy approaches the classical escape energy, providing a reference for comparison with the parametrically modulated system.
    }
    \label{fig1}
\end{figure}
\\
Figure \ref{fig1} shows the Poincaré section ($y=0$, $p_y>0$) of the unperturbed Hénon–Heiles system plotted for three different energies below the escape threshold. The phase space structure exhibits a transition from regular motion at relatively low energies to chaotic motion as the energy approaches the threshold $E = \frac{1}{6}$. This figure serves as a baseline for assessing the effects of parametric excitations later in the study.\\
\\
Recent studies have significantly advanced our understanding of the system and its generalized versions, particularly in relation to integrability and practical applications. The influence of small perturbations on the system’s integrability and the resulting chaotic behavior is examined in \cite{Gcon}. At low energy levels, asymptotic constants of motion that remain valid over extended time periods are discussed in \cite{Cos}. Techniques for controlling chaotic scattering and reducing unpredictability in randomly driven Hénon systems are explored in \cite{Coc}. A Painlevé analysis is used in \cite{Uma} to identify specific parametric conditions under which a damped Hénon system becomes integrable. The existence of multiple periodic orbits within the system is demonstrated using averaging theory in \cite{Saw}. In a more advanced study, machine learning algorithms are applied to reconstruct the chaotic dynamics of the Hénon–Heiles system \cite{Esc}, enabling accurate modeling of its behavior through data-driven methods. Additionally, the impact of higher-order nonlinearities is investigated using second-order averaging theory in \cite{Nan}.

The dynamics of the Hénon–Heiles system under external perturbations have been extensively studied. 
Systematic numerical investigations have characterized escape basins and chaotic scattering in the 
autonomous case \cite{Vallejo2025}. More recent studies have explored the effects of additive or rotating periodic forcing on scattering and escape rates \cite{EscobarRuiz2024}. Analytical approaches based on Melnikov theory and normal-form analysis have been employed to detect homoclinic splitting and stochastic layers in generalized Hénon–Heiles variants \cite{Yagasaki2010}. Furthermore, modern computational and data-driven techniques continue to use the Hénon–Heiles model as a benchmark for long-orbit classification and chaos detection. Similar dynamics have been thoroughly studied in nonlinear oscillators like the Duffing equation, where periodic driving and parametric modulation produce chaotic attractors alongside intricate bifurcation patterns~\cite{holmes_duffing_1979,moon_holmes_1979}. Comparable effects emerge in nonlinear chains and coupled-mode systems, including the Fermi--Pasta--Ulam problem, where inherent nonlinear couplings dictate energy redistribution and equipartition~\cite{ford_fpu_1992}.

Although autonomous and additively driven Hénon--Heiles systems have received considerable attention, the impact of parametrically modulating nonlinear interaction has received comparatively less attention. Whereas traditional parametric resonance theory emphasizes modulation of linear parameters like stiffness or frequency, varying nonlinear coupling coefficients fundamentally reshapes mode interactions and internal resonances. This difference proves essential in multi-degree-of-freedom Hamiltonian systems, where chaos emerges from nonlinear couplings instead of external drives. The ensuing dynamics align with the wider context of quasi-periodic bifurcations and resonant Hamiltonian frameworks~\cite{broer_quasiperiodic_2007}. To our knowledge, no prior analytical studies have systematically addressed multiplicative parametric modulation of the nonlinear coefficient in the Hénon--Heiles system.

In this work, we demonstrate that parametric modulation of the nonlinear coupling in the Hénon--Heides system modifies mode interactions. Canonical perturbation theory yields reduced normal forms that predict resonance tongues with widths scaling as $\sqrt{\varepsilon}$. Melnikov analysis establishes the threshold for separatrix splitting and chaos onset, while numerical simulations confirm the transition from localized resonances to global chaotic transport through resonance overlap. In the high-frequency limit, averaging theory uncovers an effective stiffening of the potential that suppresses chaotic dynamics. Unlike additive forcing, which primarily shifts energy levels, this mechanism emphasizes internal mode coupling and can promote combination resonances. This induces combination resonances and altered scaling laws absent in externally driven variants, enabling internal mode-coupling structures that additive forcing cannot produce.

\section{Mathematical Model Formulation}

The system under consideration is governed by the following time-dependent Hamiltonian:
\begin{equation}
\mathcal{H}(x, y, p_x, p_y, t)
= \frac{1}{2}\left(p_x^2 + p_y^2\right)
+ \frac{1}{2}\left(x^2 + y^2\right)
+ \lambda(t)\left(x^2 y - \frac{y^3}{3}\right),
\end{equation}
where the coupling parameter varies periodically as
\begin{equation}
\lambda(t) = \lambda_0 \left[ 1 + \epsilon \cos(\Omega t) \right].
\end{equation}
The time-dependent nonlinear coefficient represents a parametric modulation of the Hénon–Heiles potential, corresponding physically to a periodic deformation or “breathing” of the potential energy surface. Unlike ordinary periodic driving, which injects energy directly, parametric modulation excites the system indirectly by varying its internal stiffness and coupling strength in time. When the modulation frequency 
$\Omega$ matches one or more combinations of the system’s intrinsic frequencies, parametric resonance occurs, leading to alternating regimes of stability and chaos. For large $\Omega$, the rapid modulation averages out, effectively stiffening the potential and producing dynamic stabilization analogous to the Kapitza effect. The modulation preserves all spatial symmetries of the Hénon–Heiles potential but breaks continuous time-translation invariance and substituting it with discrete Floquet periodicity typical of Hamiltonian systems featuring periodic coefficients~\cite{yakubovich_floquet_1975}. Consequently, energy is no longer conserved in the reduced system, although the dynamics remain Hamiltonian in the extended phase space, enabling a fully symplectic perturbation analysis of resonance and chaotic behavior.

The quadratic terms represent an isotropic two-dimensional harmonic oscillator with natural frequencies 
$\omega_1$ and $\omega_2$. Hence, the system retains an internal $1{:}1$ resonance structure. 

To make the resonance properties explicit, we introduce the canonical action–angle variables corresponding 
to the $1{:}1$ oscillator:
\begin{align}
x &= \sqrt{2J_1}\cos\phi_1, & p_x &= -\sqrt{2J_1}\sin\phi_1, \nonumber \\
y &= \sqrt{2J_2}\cos\phi_2, & p_y &= -\sqrt{2J_2}\sin\phi_2.
\end{align}
In these coordinates, the unperturbed frequencies become explicit, facilitating the identification of 
resonant interactions and setting the stage for the application of canonical perturbation theory.

With the above canonical transformation, the unperturbed part of the Hamiltonian takes the simple form
\begin{equation}
\mathcal{H}_0(J) = J_1 + J_2,
\end{equation}
which corresponds to the energy of the isotropic $1{:}1$ harmonic oscillator.

Expressing the perturbation in terms of the action--angle variables, we obtain
\begin{equation}
\begin{aligned}
\mathcal{V}(\phi, J) = \sqrt{2J_2} \Bigg[ &
\left(J_1 - \tfrac{J_2}{2}\right)\cos\phi_2 
+ \tfrac{J_1}{2} \big( \cos(2\phi_1 + \phi_2) + \cos(2\phi_1 - \phi_2) \big)
- \tfrac{J_2}{6}\cos(3\phi_2) \Bigg].
\end{aligned}
\end{equation}

From this expansion, it follows that the only angle combinations capable of resonant interaction are associated with the following mode vectors:
\[
k = (0, 1),\ (2, \pm 1),\ (0, 3).
\]
These correspond to the fundamental, second-order, and third-harmonic resonances of the system, respectively.

To render the time-dependent Hamiltonian autonomous, we promote the temporal variable $t$ to a genuine 
angle variable $\theta$, with conjugate action $J_{\theta}$. In this extended phase space, the Hamiltonian 
becomes time independent and the system evolves in a higher-dimensional symplectic manifold. 
This extension restores energy conservation and reveals resonant structures through integer relations among 
the frequencies $(1,\,1,\,\Omega)$.

The resulting autonomous Hamiltonian is expressed as
\begin{equation}
\mathcal{H}(J_1, J_2, \phi_1, \phi_2, \theta, J_{\theta})
= \mathcal{H}_0(J)
+ \lambda_0 \mathcal{V}(\phi, J)
+ \epsilon \lambda_0 \cos\theta\, \mathcal{V}(\phi, J)
+ \Omega J_{\theta}.
\end{equation} 

We express the cubic perturbation as a finite Fourier series
\begin{equation}
\mathcal{V}(\phi,J)
= \sum_{k \in \kappa} \mathcal{V}_{k}(J)\,\cos\!\big(k\!\cdot\!\phi\big),
\qquad 
\kappa=\{(0,1),(2,1),(2,-1),(0,3)\},
\end{equation}
where $k\!\cdot\!\phi := k_1\phi_1 + k_2\phi_2$. The (exact) mode amplitudes are
\begin{align}
\mathcal{V}_{(0,1)}(J) &= \sqrt{2J_2}\!\left(J_1 - \tfrac{J_2}{2}\right), \\
\mathcal{V}_{(2,\pm 1)}(J) &= \tfrac{1}{2}\sqrt{2J_2}\,J_1, \\
\mathcal{V}_{(0,3)}(J) &= -\,\tfrac{1}{6}\sqrt{2J_2}\,J_2 .
\end{align}

With this notation, the autonomous, extended Hamiltonian can be written compactly as
\begin{equation}
\mathcal{H}(J,\phi,\theta,J_{\theta})
= \underbrace{(J_1+J_2)}_{\mathcal{H}_0(J)}
+ \lambda_0\!\left(1+\epsilon \cos\theta\right)\!
\sum_{k\in\kappa}\mathcal{V}_{k}(J)\,\cos\!\big(k\!\cdot\!\phi\big)
+ \Omega J_{\theta}.
\end{equation}
The corresponding equations of motions are as follows
\paragraph{Angle variables:}
\begin{equation}
\dot{\phi}_i
= \frac{\partial \mathcal{H}}{\partial J_i}
= 1
+ \lambda_0 (1+\epsilon \cos\theta)
  \sum_{k\in\kappa}
  \frac{\partial \mathcal{V}_k}{\partial J_i}(J)
  \cos(k\!\cdot\!\phi),
\qquad i = 1,2.
\end{equation}

\paragraph{Action variables:}
\begin{equation}
\dot{J}_i
= -\,\frac{\partial \mathcal{H}}{\partial \phi_i}
= \lambda_0 (1+\epsilon \cos\theta)
  \sum_{k\in\kappa}
  k_i\,\mathcal{V}_k(J)
  \sin(k\!\cdot\!\phi),
\qquad i = 1,2.
\end{equation}

\paragraph{Drive variables:}
\begin{equation}
\dot{\theta} = \frac{\partial \mathcal{H}}{\partial J_{\theta}} = \Omega,
\qquad
\dot{J}_{\theta}
= -\,\frac{\partial \mathcal{H}}{\partial \theta}
= \epsilon\,\lambda_0\,\sin\theta
  \sum_{k\in\kappa}
  \mathcal{V}_k(J)\cos(k\!\cdot\!\phi).
\end{equation}

\section{Weak perturbation Theory(canonical; $\epsilon<<1$)}

\subsection{Lie--Deprit Normalization at First Order}

The autonomous Hamiltonian is not integrable due to the presence of the oscillatory term  $\cos\theta\,\mathcal{V}(\phi,J)$. Therefore, the next goal is to find a canonical transformation $(J,\phi,\theta,J_{\theta}) \rightarrow (J',\phi',\theta',J_{\theta}')$ that simplifies the Hamiltonian, 
in particular by eliminating all non-resonant oscillations to first order in $(\lambda_0,\,\epsilon\lambda_0)$. This is achieved through canonical (Lie–Deprit) perturbation theory. We seek a near-identity canonical transformation generated by the function $\mathcal{W}_1(\phi, \theta, J)$, which is small and of order $\lambda_0$ or $\epsilon\lambda_0$, 
such that the transformed Hamiltonian becomes
\begin{equation}
\mathcal{H}' = e^{L_{\mathcal{W}_1}}\mathcal{H} 
= \mathcal{H} + \{\mathcal{W}_1, \mathcal{H}\} 
+ \frac{1}{2}\{\mathcal{W}_1, \{\mathcal{W}_1, \mathcal{H}\}\} + \cdots,
\end{equation}
and contains no terms oscillating with fast angular combinations. 
Here, $L_{\mathcal{W}_1}$ denotes the Lie operator generated by $\mathcal{W}_1$, defined as 
$L_{\mathcal{W}_1}(\cdot) = \{\mathcal{W}_1, (\cdot)\}$.

We decompose the Hamiltonian into an integrable part $\mathcal{H}_0'$ and an angle-dependent perturbation 
part $\mathcal{H}_1$, expressed as
\begin{equation}
\begin{aligned}
\mathcal{H}_0' &= \mathcal{H}_0(J) + \Omega J_{\theta}, \\
\mathcal{H}_1  &= \lambda_0\,\mathcal{V}(\phi, J) 
                + \epsilon\,\lambda_0\,\cos\theta\,\mathcal{V}(\phi, J).
\end{aligned}
\end{equation}
Here, $\mathcal{H}_0'$ represents the integrable Hamiltonian, while $\mathcal{H}_1$ contains the 
angle-dependent perturbative terms. We choose $\mathcal{W}_1$ such that it cancels the unwanted fast (non-resonant) components of 
$\mathcal{H}_1$. To first order, the transformed Hamiltonian can be written as
\begin{equation}
\mathcal{H}' = \mathcal{H}_0 + \mathcal{H}_1 + \{\mathcal{W}_1, \mathcal{H}_0\} + \cdots,
\end{equation}
which leads to the homological equation
\begin{equation}
\{\mathcal{W}_1, \mathcal{H}_0\} = -(\mathcal{H}_1)_{\text{non-res}}.
\end{equation}
Thus, $\mathcal{W}_1$ serves as the generating function that removes the fast oscillatory (non-resonant) terms up to first order. The solution to the homological equation is obtained by expanding the perturbative part of the Hamiltonian in a Fourier series as
\begin{equation}
\mathcal{H}_1 = \sum_{k,m} \mathcal{H}_{1;k,m}(J)\,\cos\!\big(k\!\cdot\!\phi + m\theta\big),
\end{equation}
where $k = (k_1, k_2) \in \mathbb{Z}^2$ and $m \in \mathbb{Z}$. The Fourier coefficients are given by
\begin{equation}
\begin{aligned}
\mathcal{H}_{1;k,0} &= \lambda_0 \mathcal{V}_k(J)\\
\mathcal{H}_{1;k,\pm1} &= \frac{\epsilon \lambda_0}{2}\mathcal{V}_k(J)
\end{aligned}
\end{equation}
Now look at the Possion bracket 
\begin{equation}
\begin{aligned}
\lbrace \mathcal{W}_1, \mathcal{H}_0  \rbrace &= \sum_i \frac{\partial \mathcal{W}_1 }{\partial \phi_i}\frac{\partial \mathcal{H}_0 }{\partial J_i} +\frac{\partial \mathcal{W}_1 }{\partial \theta}\frac{\partial \mathcal{H}_0 }{\partial J_\theta}\\&=
\sum_i \omega_1 \frac{\partial \mathcal{W}_1 }{\partial \phi_i} + \Omega \frac{\partial \mathcal{W}_1 }{\partial \theta}
\end{aligned}
\end{equation}
Assume 
\begin{equation}
\mathcal{W}_1 = \sum_{k,m} \mathcal{W}_{1; k,m}(J) sin(k.\phi+m\phi)
\end{equation}

The Poisson bracket between $\mathcal{W}_1$ and $\mathcal{H}_0$ is given by
\begin{equation}
\{\mathcal{W}_1, \mathcal{H}_0\}
= \sum_{k,m} (k\!\cdot\!\omega + m\Omega)\,
\mathcal{W}_{1;k,m}(J)\,\cos\!\big(k\!\cdot\!\phi + m\theta\big).
\end{equation}
This implies that
\begin{equation}
(k\!\cdot\!\omega + m\Omega)\,\mathcal{W}_{1;k,m}(J)
= -\,\mathcal{H}_{1;k,m}(J),
\end{equation}
for non-resonant terms, yielding the explicit expression
\begin{equation}
\mathcal{W}_{1;k,m}(J)
= \frac{\mathcal{H}_{1;k,m}(J)}{(k\!\cdot\!\omega + m\Omega)}\,
\sin\!\big(k\!\cdot\!\phi + m\theta\big).
\end{equation}

If any denominator $(k\!\cdot\!\omega + m\Omega) \approx 0$ (resonance), the corresponding term cannot be 
removed since it would make $\mathcal{W}_1$ divergent. Such resonant terms are retained in the transformed 
Hamiltonian $\mathcal{H}'$. Hence, after applying the canonical transformation generated by $\mathcal{W}_1$, 
the new Hamiltonian takes the form
\begin{equation}
\mathcal{H}' = \mathcal{H}_0 + \mathcal{H}_{\mathrm{res}} 
+ \mathcal{O}(\lambda_0^2,\, \epsilon^2).
\end{equation}
$\mathcal{H}_{\mathrm{res}}$ consists only of the resonant harmonics. The resonant part of the Hamiltonian is given by

\begin{equation}
\mathcal{H}_{\mathrm{res}} 
= \sum_{(k,m)_{\mathrm{res}}}
\mathcal{H}_{1;k,m}(J)\,\cos\!\big(k\!\cdot\!\phi + m\theta\big).
\end{equation}

For $k = (0,1)$ and $m = -1$, resonance occurs when $1 - \Omega \approx 0$, corresponding to the 
primary resonance tongue. For $k = (0,3),\, m = -1$ or $k = (2,\pm1),\, m = -1$, resonance arises 
when $3 - \Omega \approx 0$, corresponding to the higher-order tongues.  The reduced Hamiltonian thus obtained can be transformed into an effective one-degree-of-freedom  pendulum model, from which the tongue boundaries and separatrix widths can be derived.

\begin{figure*}[t]
\centering
\includegraphics[width=0.5\linewidth]{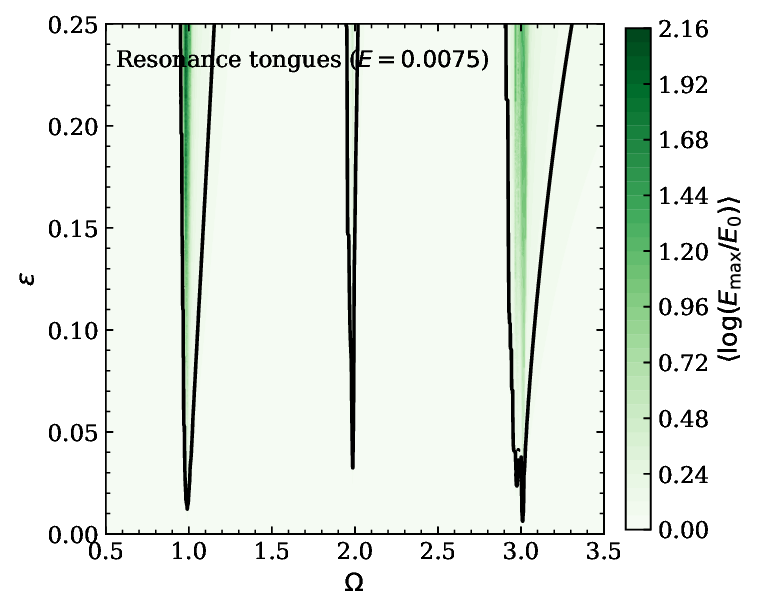}
\caption{
{\bf Resonance tongues} in the parametrically modulated Hénon--Heiles system at low energy ($E \approx 0.0075$). Narrow instability bands emerge at $\Omega \approx 1$, $\Omega \approx 2$, and $\Omega \approx 3$, corresponding to the primary, combination ($k = (2, -1)$), and higher-harmonic resonances predicted by normal-form analysis. Tongue widths increase with modulation amplitude $\epsilon$, consistent with perturbative scaling laws.
}
\label{fig2}
\end{figure*}

At sufficiently low energies ($E = 0.0075$), the Hénon--Heiles system operates in a weakly nonlinear regime where deviations from the underlying harmonic oscillator remain small, preserving near-integrable motion. Here, the canonical perturbation framework and associated normal-form reduction accurately describe the dynamics, with numerical resonance tongues providing a direct test of the analytical theory. Distinct instability bands emerge at $\Omega \approx 1$, $\Omega \approx 2$, and $\Omega \approx 3$, matching the fundamental, combination, and higher-order resonances from the condition $k \cdot \omega + m \Omega \approx 0$. Tongue widths scale with modulation amplitude $\epsilon$ as expected perturbatively, confirming weak nonlinear interactions dominate. Notably, the $\Omega \approx 2$ resonance demonstrates that even modest nonlinearities activate higher-order coupling channels, enabling combination resonances beyond primary and harmonic modes [see FIG.~\ref{fig2}].

If we had continued with the original time-dependent Hamiltonian $\mathcal{H}(x,p,t)$, the Poisson bracket 
$\{\mathcal{W}_1, \mathcal{H}_0\}$ would not contain the $m\Omega$ terms, since time is not a canonical variable. 
By introducing the pair $(\theta, J_{\theta})$, time becomes a canonical angle, and its conjugate frequency 
$\Omega$ naturally appears in the denominators $(k\!\cdot\!\omega + m\Omega)$. This is the fundamental reason for rendering the system autonomous: it allows the time modulation to be treated as an additional periodic degree of freedom, thereby enabling the direct application of standard canonical perturbation theory.
 \subsubsection{Reduction to Pendulum Normal Form Near $\Omega\approx1$}
By retaining only the resonant harmonic proportional to $\cos(\phi_2 - \theta)$, we introduce the slow variables
\begin{equation}
\Psi = \phi_2 - \theta, \qquad I = J_2,
\end{equation}
and define the corresponding amplitude as
\begin{equation}
A(J_1, I) = \sqrt{2I}\!\left(J_1 - \tfrac{I}{2}\right).
\end{equation}

This leads to a reduced Hamiltonian that captures the slow dynamics near the resonance,
\begin{equation}
\mathcal{H}_{\mathrm{NF}} 
= \Delta I + \frac{\epsilon\,\lambda_0}{2}\,A(J_1, I)\,\cos\Psi,
\end{equation}
where $\Delta = 1 - \Omega$ is the detuning parameter quantifying how far the drive frequency is from exact resonance.

For each value of $\Omega$, there exists an exact resonant action $I_0$ satisfying
\begin{equation}
k\!\cdot\!\omega(I_0) = \Omega,
\end{equation}
which, in this case, implies $I = \Omega$ at $I = I_0$. For small detuning, when $\Omega$ is close to the natural frequency, the actual resonant trajectory deviates slightly from $I_0$. Hence, the natural frequency is expanded in the vicinity of this resonance point.

We expand the frequency near the resonant action $I_0$ as
\begin{equation}
k\!\cdot\!\omega(I) \approx 1 + \alpha(I_0)\,(I - I_0), 
\qquad 
\alpha(I_0) = \left.\frac{\partial (k\!\cdot\!\omega)}{\partial I}\right|_{I_0},
\end{equation}
where $\alpha(I_0)$ denotes the local frequency shear, characterizing how the natural frequency varies 
with amplitude due to nonlinear effects. 

Shifting the origin of the action variable as $I \rightarrow I - I_0$, neglecting constant terms, 
and defining $a = \alpha(I_0)$, the Hamiltonian reduces to
\begin{equation}
\mathcal{K}(I, \Psi) = \frac{a}{2}I^2 + B\,\cos\Psi, 
\qquad 
B = \frac{\epsilon\,\lambda_0}{2}\,A(J_1, I_0).
\end{equation}

This Hamiltonian is mathematically identical to that of a simple pendulum, where $I$ plays the role of the 
angular momentum and $\Psi$ corresponds to the pendulum angle. 

\begin{itemize}
\item[(1)] The first term, $\tfrac{a}{2}I^2$, represents the kinetic energy, quadratic in the deviation from resonance.
\item[(2)] The second term, $B\,\cos\Psi$, corresponds to the potential energy associated with the slow oscillations.
\end{itemize}

The pendulum Hamiltonian possesses a separatrix that distinguishes between trapped (librational) and 
untrapped (rotational) motion. At the separatrix, the total energy equals $E_{\mathrm{sep}} = |B|$. 
The half-width of this separatrix in the action ($I$) space is given by
\begin{equation}
\Delta I_{\mathrm{sep}} 
= 4\sqrt{\frac{|B|}{|a|}} 
= 4\sqrt{\frac{|\epsilon\,\lambda_0|}{2|a|}\,|A(J_1, I_0)|} 
\propto \sqrt{\epsilon}.
\end{equation}
This result indicates that the width of the resonance island scales as $\sqrt{\epsilon}$; doubling the 
modulation amplitude increases the island width by approximately 40\%. Such scaling is universal for 
weakly driven Hamiltonian resonances.

In the pendulum analogy, motion is trapped (corresponding to instability) when the effective energy 
lies within the separatrix region. This occurs when the detuning $\Delta = 1 - \Omega$ is sufficiently 
small, such that the term $\Delta I$ does not destroy libration. Mathematically, the boundary is reached 
when the effective oscillation frequency of the slow motion vanishes, i.e.,
\begin{equation}
|\Omega - 1| \approx c\,\epsilon\,\lambda_0,
\end{equation}
where the proportionality constant $c$ is obtained by matching the slope of the effective potential, yielding
\begin{equation}
c = \frac{1}{2}\left|\frac{A(J_1, I_0)}{\sqrt{|\alpha(I_0)|}}\right|.
\end{equation}

These expressions quantitatively determine how the parametric instability shifts and how the resonant 
island extends in phase space. They define the linear bounding law for the resonance tongue in the 
$(\Omega, \epsilon)$ plane. Inside the tongue, the slow variable $\Psi$ librates, energy is periodically pumped into the system, and parametric instability or phase locking occurs. Outside the tongue, $\Psi$ rotates freely, the modulation averages out, and the system exhibits quasi-periodic behavior. The tongue boundary marks the resonance threshold—beyond it, small oscillations either remain bounded or grow exponentially, signifying the onset of parametric resonance.
\subsubsection{Combination resonances $\Omega\approx3$ from $k=(0,3)$ and $k=(2,\pm1)$}
Define 
\begin{equation}
\begin{aligned}
A_3(I) &= -\sqrt{2I}\frac{I}{6}\\
A_{\pm} (J_1,I) &= \sqrt{2I}\frac{J_1}{2}
\end{aligned}
\end{equation}
Retaining the terms $\cos\theta\,\cos3\phi_2$ and $\cos\theta\,\cos(2\phi_1 \pm \phi_2)$ introduces 
the slow angles $\Psi_3 = 3\phi_2 - \theta$ and $\Psi_{\pm} = 2\phi_1 \pm \phi_2 - \theta$, with the 
corresponding detunings $\Delta_3 = 3 - \Omega$ and $\Delta_{\pm} = 1 - \Omega$ for the “$-$” branch. 
The associated reduced Hamiltonians take the same pendulum form as before, with the coupling amplitude 
$B$ replaced by $\tfrac{\epsilon\,\lambda_0}{2}A_3$ or $\tfrac{\epsilon\,\lambda_0}{2}A_{\pm}$, and the 
detuning $\Delta$ replaced by $\Delta_3$ or $\Delta_{\pm}$, respectively.

\begin{figure}[t]
    \centering
    \includegraphics[width=0.65\linewidth]{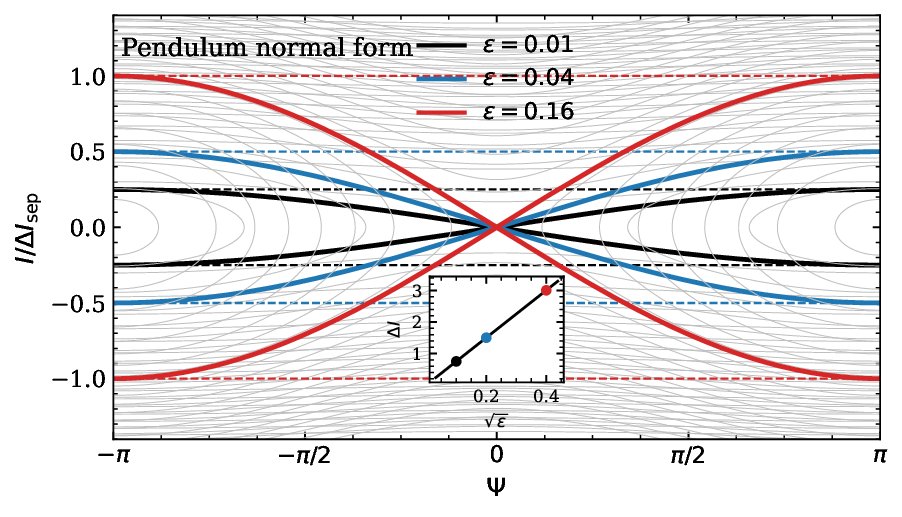}
    \caption{
    \textbf{Pendulum normal-form phase portrait near the primary $1\!:\!1$ resonance.}
The contours show level sets of the reduced Hamiltonian
    \(
        K(I,\Psi)=\tfrac{a}{2}I^{2}+B\cos\Psi,
        \quad
        B=\tfrac{\varepsilon\lambda_{0}}{2}A(J_{1},I_{0}),
    \)
with the vertical axis normalized by the analytically predicted separatrix half-width
    \(
        \Delta I_{\rm sep}=2\sqrt{B/|a|}.
    \)
    For each drive amplitude $\varepsilon\in\{0.01,0.04,0.16\}$, the thick colored curve denotes the separatrix $K=|B|$, while the dashed horizontal lines show the corresponding analytical half-width 
    \(
        \Delta I_{\rm sep}/\Delta I_{\rm sep}(0.16).
    \)
    The collapse of the normalized separatrices demonstrates excellent agreement with the pendulum normal form. The inset displays the predicted scaling $\Delta I_{\rm sep}\propto\sqrt{\varepsilon}$: the numerically measured values lie directly on the analytical line.
    }
    \label{fig3}
\end{figure}

From Figure~\ref{fig3}, the numerically obtained resonance structure compares well with the pendulum normal form predictions. Extracting \( A(J_1, I_0) \) and \(\alpha(I_0)\) from the unperturbed dynamics, the reduced Hamiltonian yields a separatrix with half-width scaling as \(\Delta I_{\rm sep} = 2 \sqrt{B/|a|} \propto \sqrt{\varepsilon}\). Plotting the phase portrait with the dimensionless coordinate \( I / \Delta I_{\rm sep}(0.16) \), separatrices for \(\varepsilon = 0.01, 0.04, 0.16\) collapse onto the universal pendulum shape. The inset confirms the \(\Delta I_{\rm sep} \propto \sqrt{\varepsilon}\) scaling, with all points on the analytic curve. This validates the pendulum approximation in capturing the resonance island geometry and scaling in the Hénon–Heiles system.

This reduction establishes that the parametrically driven Hénon–Heiles system behaves locally as a nonlinear pendulum near resonance, providing a direct analytical explanation for the resonance tongues observed numerically.
\subsection{Melnikov Analysis}
The unperturbed Hénon–Heiles Hamiltonian is conservative, with fixed energy surfaces that possess 
homoclinic orbits at the escape energy. The homoclinic trajectory $z_0(t)$ connects a saddle point to itself. 
In phase space, the stable and unstable manifolds of the saddle coincide, forming a homoclinic loop. When a time-periodic perturbation is introduced, the separatrix is slightly modulated. The key question is whether the stable and unstable manifolds of the perturbed system continue to coincide (indicating regular  motion) or whether they split (leading to chaotic dynamics or escape). This issue is traditionally tackled via the Melnikov technique, which yields a leading-order quantification of separatrix splitting in periodically perturbed Hamiltonian systems~\cite{holmes_marsden_1982}.

The Melnikov function $M(t_0)$ measures the first-order distance in phase space between the perturbed 
stable and unstable manifolds of the saddle and is defined as
\begin{equation}
M(t_0) = \int_{-\infty}^{+\infty} 
\!dt\, 
\big\{\mathcal{H}_{\mathrm{aut}}, 
\, \epsilon\,\lambda_0\,\cos[\Omega(t + t_0)]\,\mathcal{V}\big\}\big|_{z_0(t)},
\end{equation}
where $\{\cdot,\cdot\}$ denotes the Poisson bracket, $t_0$ is the phase shift of the periodic drive, 
and $\mathcal{H}_{\mathrm{aut}}$ generates the unperturbed motion. Simple zeros in the Melnikov function signal transverse crossings of invariant manifolds and the emergence of chaotic motion, as demonstrated in foundational studies of Hamiltonian chaos~\cite{holmes_marsden_1982}.

We have
\begin{equation}
\{\mathcal{H}_{\mathrm{aut}}, \mathcal{V}\} = \dot{\mathcal{V}}(z_0, t).
\end{equation}
Hence, the Melnikov function can be written as
\begin{equation}
M(t_0) = \epsilon\,\lambda_0 
\int_{-\infty}^{+\infty} \!dt\, 
\dot{\mathcal{V}}(z_0(t))\,\cos[\Omega(t + t_0)].
\end{equation}

Integrating by parts, the boundary term vanishes since $z_0(t) \rightarrow$ saddle as 
$t \rightarrow \pm\infty$. Therefore,
\begin{equation}
\begin{aligned}
M(t_0) 
&= \epsilon\,\lambda_0\,\Omega 
\int_{-\infty}^{+\infty} \!dt\, 
\mathcal{V}(z_0(t))\,\sin[\Omega(t + t_0)] \\
&= \epsilon\,\lambda_0\,\Omega 
\big[A(\Omega)\,\sin(\Omega t_0) + B(\Omega)\,\cos(\Omega t_0)\big],
\end{aligned}
\end{equation}
where 
\begin{equation}
A(\Omega) = \int_{-\infty}^{+\infty} \!dt\, \mathcal{V}(z_0(t))\,\cos(\Omega t), 
\qquad
B(\Omega) = \int_{-\infty}^{+\infty} \!dt\, \mathcal{V}(z_0(t))\,\sin(\Omega t).
\end{equation}

By defining the complex Fourier transform as
\begin{equation}
\hat{\mathcal{V}}(\Omega) 
= \int_{-\infty}^{+\infty} 
\mathcal{V}(z_0(t))\,e^{-i\Omega t}\,dt,
\end{equation}
the coefficients in the Melnikov function can be expressed in terms of its real and imaginary parts as
\begin{equation}
A(\Omega) = \Re\!\big[\hat{\mathcal{V}}(\Omega)\big], 
\qquad 
B(\Omega) = \Im\!\big[\hat{\mathcal{V}}(\Omega)\big].
\end{equation}

The amplitude of the Melnikov function is then given by
\begin{equation}
|M_{\mathrm{max}}| = \epsilon\,\lambda_0\,\Omega\,|\hat{\mathcal{V}}(\Omega)|,
\end{equation}
where $\hat{\mathcal{V}}(\Omega)$ is the Fourier transform of $\mathcal{V}(z_0(t))$ 
evaluated along the separatrix. 

Thus, $M(t_0)$ quantifies the energy exchange per cycle between the external modulation and the separatrix orbit. When this overlap is sufficiently large, the separatrix breaks up, leading to chaotic scattering and the onset of global chaos.
\subsubsection{Threshold Condition-Onset of Chaos/Escape}

For chaos(transverse intersection) to occur $M(t_0)$ must have at least one zero (change of sign)
\begin{equation}
\exists \hspace{.1cm} t_0 \hspace{.1cm} s.t\hspace{.1cm} M(t_0)=0
\end{equation}
Simple zeros of $M(t_0)$ occur when its amplitude is non-zero, implying transverse intersections of invariant manifolds. The magnitude of $M(t_0)$ governs the strength of separatrix splitting and thus the degree of chaotic transport. Since $M(t_0)$ oscillates sinusoidally, such zeros arise whenever its amplitude exceeds zero. We define $\epsilon_c(\Omega)$ as the smallest $\epsilon$ for which this splitting enables escape, marking the onset of global chaotic transport.

\begin{equation}
\epsilon_c(\Omega) \propto \frac{1}{|\lambda_0| \Omega |\hat{V}(\Omega)|}
\end{equation}
This criterion is universal and does not depend on the specific system details—any periodically driven 
Hamiltonian system with a separatrix exhibits the same scaling behavior. In the Hénon–Heiles system, it 
quantifies the transition from local chaotic diffusion to global escape. When the stable and unstable 
manifolds intersect transversely, the local dynamics near the saddle point form a Smale horseshoe—an 
infinite set of nested lobes corresponding to chaotic scattering trajectories. 

\begin{figure*}[t]
\centering
\includegraphics[width=0.5\linewidth]{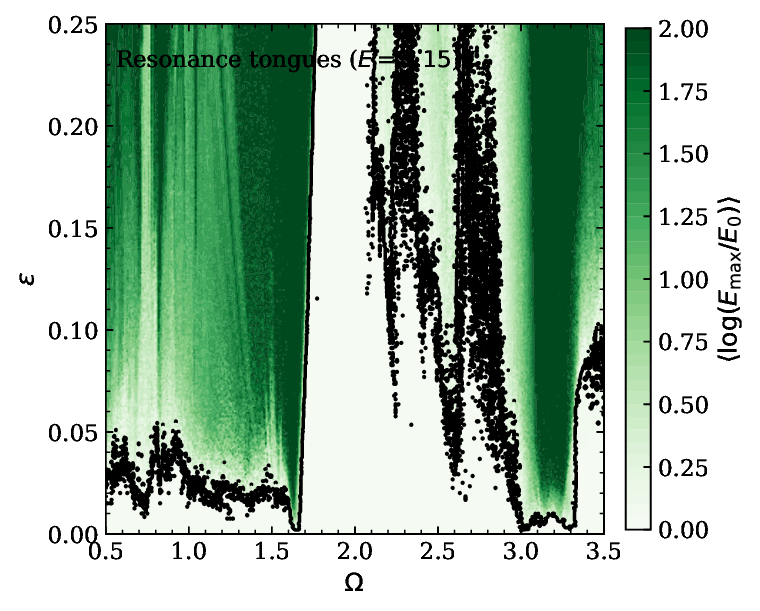}
\caption{
{\bf Resonance tongues} in parametrically modulated Hénon--Heiles system at $E \approx 0.15$. Narrow instability bands at $\Omega \approx 1, 2, 3$ correspond to primary, combination ($k = (2, -1)$), and higher-harmonic resonances from normal-form analysis; widths increase with $\epsilon$ per perturbative scaling. At larger $\epsilon$, bands broaden/overlap (Chirikov criterion), with boundary scattering indicating separatrix splitting and stochastic layers (Melnikov), tracing perturbative-to-chaotic transition.
}
\label{fig4}
\end{figure*}

The color map of $\log(E_{\max}/E_0)$ quantifies energy growth under parametric modulation, revealing the full resonant and chaotic structure. At small $\epsilon$, narrow vertical bands of enhanced growth appear near $\Omega \approx 1$, $\Omega \approx 2$, and $\Omega \approx 3$, matching the primary, combination, and higher-harmonic resonances from perturbative analysis[FIG.~\ref{fig2}]. Normal-form theory accurately describes this localized growth regime. As $\epsilon$ increases, bands broaden and overlap, creating extended regions of strong amplification---signaling resonance interactions and chaotic transport per the Chirikov overlap criterion. Scattered structures near resonance boundaries indicate separatrix splitting and stochastic layers, consistent with Melnikov analysis. Thus, the figure traces the transition from perturbative, weakly nonlinear dynamics to global chaos driven by parametric modulation[see FIG.~\ref{fig4}].
To illustrate the phase-space effects of parametric resonances, we examine Poincar\'e sections at representative driving frequencies. These sections visualize the transition from regular to chaotic motion as resonance conditions are approached.
\begin{figure*}[t]
\centering
\includegraphics[width=0.5\linewidth]{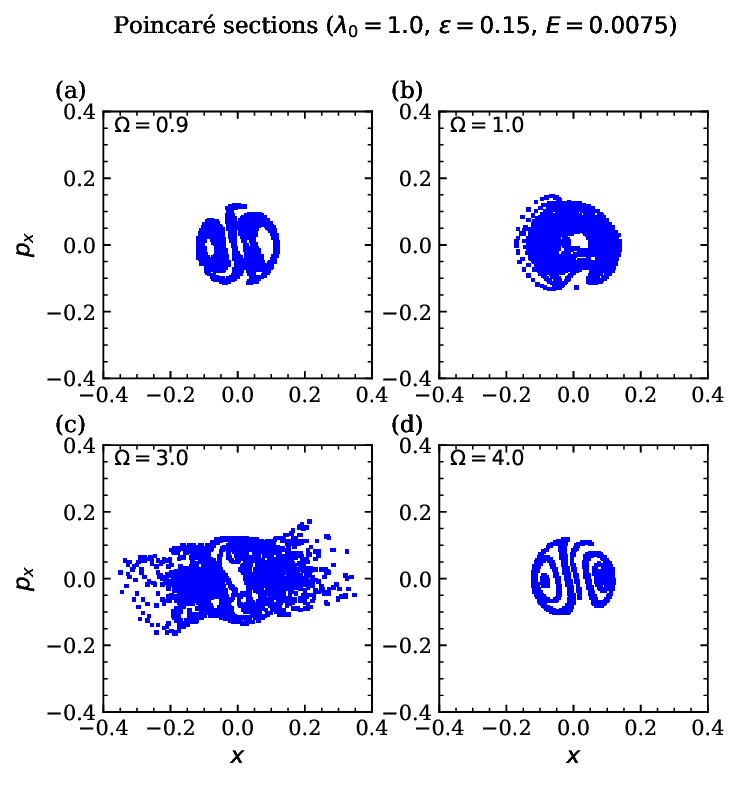}
\caption{
Poincar\'e sections of the parametrically modulated H\'enon--Heiles system 
at fixed energy ($E = 0.0075$) and modulation amplitude $\epsilon = 0.15$ 
for different driving frequencies $\Omega$. 
(a) $\Omega = 0.9$: off-resonant case showing regular motion. 
(b) $\Omega = 1.0$: primary resonance with onset of stochasticity. 
(c) $\Omega = 3.0$: strong resonance with pronounced chaotic spreading. 
(d) $\Omega = 4.0$: off-resonant regime with restored regular dynamics(Kapitza effect). 
}
\label{fig5}
\end{figure*}
Consequently, even a very small modulation amplitude $\epsilon$ can induce chaotic ejection of orbits 
from the potential well. The amplitude $|\hat{\mathcal{V}}(\Omega)|$ determines how effectively the external modulation couples to the separatrix motion; it typically exhibits a band-pass dependence, 
being strongest near the natural oscillation frequency of the separatrix and weakening at higher 
driving frequencies $\Omega$. Therefore, the chaotic threshold $\epsilon_c(\Omega)$ attains its minimum 
near resonance and increases for larger $\Omega$. This establishes the Melnikov mechanism as the fundamental origin of chaos and escape in the weakly 
driven Hénon–Heiles system.
\subsection{When $\Omega >> 1$}
When the driving frequency is much higher than the system’s natural frequency, the system cannot 
follow the rapid oscillations of the external drive in real time. Instead, the motion experiences only 
the time-averaged influence of these fast modulations, effectively evolving in an averaged or 
“effective” potential. This regime is commonly referred to as the high-frequency limit or the 
Kapitza regime. This phenomenon connects directly to vibrational stabilization processes, particularly the Kapitza effect, and can be rigorously obtained through high-frequency averaging or Magnus expansion methods~\cite{kapitza_1951,blanes_magnus_2009}.

We introduce a time-periodic generating function such that the canonical transformation is defined by
\begin{equation}
\mathcal{H}_{\mathrm{eff}} = e^{L_{G_1}}\mathcal{H},
\end{equation}
where $G_1 = G_1(x, y, p_x, p_y, \theta)$. The corresponding homological equation is
\begin{equation}
\{G_1, \Omega J_{\theta}\} + \epsilon\,\lambda_0\,\cos\theta\,\mathcal{V} = 0.
\end{equation}
Since $\{G_1, \Omega J_{\theta}\} = \Omega\,\partial_{\theta} G_1$, we obtain
\begin{equation}
G_1 = \frac{\epsilon\,\lambda_0}{\Omega}\,\sin\theta\,\mathcal{V}(x, y).
\end{equation}

The effective Hamiltonian is then expressed as
\begin{equation}
\mathcal{H}_{\mathrm{eff}} 
= \mathcal{H}_{\mathrm{aut}} 
+ \frac{1}{4\Omega^{2}}\,
\{F, \{F, \mathcal{H}_{\mathrm{aut}}\}\}
+ \mathcal{O}\!\left(\frac{1}{\Omega^{3}}\right),
\end{equation}
which follows from the second-order expansion of the Lie transformation. 
\begin{equation}
\{\mathcal{V}, \{\mathcal{V}, \mathcal{U}\}\} = 0,
\end{equation}
\begin{equation}
\{\mathcal{V}, \{\mathcal{V}, \mathcal{T}\}\} 
= \sum_{i = x, y} \left(\partial_{q_i} \mathcal{V}\right)^2
= (\partial_x \mathcal{V})^2 + (\partial_y \mathcal{V})^2,
\end{equation}
\begin{equation}
\begin{aligned}
\partial_x \mathcal{V} &= 2xy, \\
\partial_y \mathcal{V} &= x^2 - y^2.
\end{aligned}
\end{equation}
We have
\begin{equation}
\{\mathcal{V}, \{\mathcal{V}, \mathcal{H}\}\} = x^4 + 2x^2y^2 + y^4,
\end{equation}
and therefore,
\begin{equation}
\mathcal{H}_{\mathrm{eff}} 
= \mathcal{H}_{\mathrm{aut}} 
+ \frac{\epsilon^2 \lambda_0^2}{4 \Omega^2}
\left(x^4 + 2x^2y^2 + y^4\right)
+ \mathcal{O}\!\left(\frac{\epsilon^3}{\Omega^3}\right).
\end{equation}

This represents the form of the effective potential in the high-frequency limit. The additional term
\begin{equation}
V_{\mathrm{eff}}(x, y)
= \frac{\epsilon^2 \lambda_0^2}{4 \Omega^2}
\left(x^4 + 2x^2y^2 + y^4\right)
\end{equation}
is positive definite and quartic in nature. It introduces a stiffening effect, causing the effective potential 
to grow more rapidly with increasing displacement. Consequently, large excursions in $(x, y)$ become 
energetically more expensive, leading to more regular motion and suppression of chaos. 

As $\Omega$ increases, the magnitude of the additional potential decreases as $1/\Omega^2$. This represents 
a nonlinear renormalization of the potential—the system effectively becomes stiffer under rapid modulation. 
The origin of the quartic correction is purely dynamical, arising from the vibrational kinetic energy associated with the fast oscillations of the coordinates under the external drive.
\subsection{Chirikov Resonance Overlap Estimate}
For resonances $r$ and $s$, centered at actions $I_r$ and $I_s$ with pendulum normal forms and associated half-widths, the Chirikov resonance-overlap criterion furnishes a global indicator for the emergence of global chaos~\cite{chirikov_1979}. The Chirikov overlap criterion is expressed as
\begin{equation}
|I_r - I_s| \leq \Delta I_r + \Delta I_s.
\end{equation}
This condition provides a global estimate for the onset of widespread chaos, yielding
\begin{equation}
\epsilon_{\mathrm{ov}} 
\sim 
\sqrt{\frac{|a|\,(I_r - I_s)^2}{8\,|\lambda_0|\,|V_k(J_0)|}}.
\end{equation}
To illustrate perturbative breakdown at higher energies, we examine Poincar\'e sections at $E = 0.15$ near the escape threshold. In resonant/near-resonant cases (panels (a)--(c)), phase-space structures distort strongly; many trajectories fail to return, yielding sparse distributions indicative of resonance-overlap-driven transport and escape. Conversely, off-resonant high-frequency driving (panel (d)) produces dense, bounded structures with repeated section returns. This underscores resonances' critical role in enabling transport, while off-resonant modulation suppresses escape.
\begin{figure*}[t]
\centering
\includegraphics[width=0.5\linewidth]{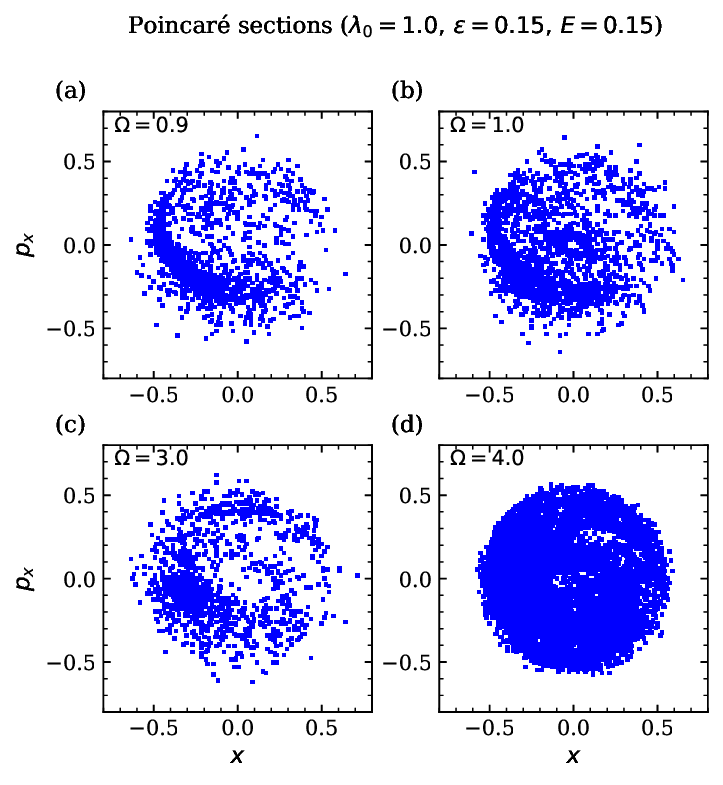}
\caption{{\bf Poincar\'e sections} of the parametrically modulated H\'enon--Heiles system at $E = 0.15$ and $\epsilon = 0.15$. 
(a)--(c) Near-resonant cases ($\Omega = 0.9, 1.0, 3.0$) show strong phase-space distortion and sparse point distributions, indicating that many trajectories undergo large excursions or escape from the sampled region due 
to resonance-driven transport. 
(d) Off-resonant case ($\Omega = 4.0$) exhibits a densely populated structure, as trajectories remain bounded and repeatedly return to the section. This contrast demonstrates that parametric resonances enhance transport and escape, while off-resonant driving suppresses these effects.
}
\label{fig6}
\end{figure*}

\section{Discussion}
Parametric modulation of nonlinear coupling reshapes the effective potential and, crucially, the internal resonance structure by directly altering mode interactions. Unlike additive forcing, which mainly shifts energy levels or linear responses, this modifies energy exchange pathways in multi-degree-of-freedom Hamiltonians, enabling combination resonances and changing resonant transfer conditions.

Normal-form reduction, Melnikov analysis, and resonance-overlap arguments unify the transition from regular to chaotic dynamics. In the weak-modulation regime, resonance tongues scale as $\sqrt{\varepsilon}$, with Melnikov integrals capturing separatrix splitting and chaotic layer onset. Higher amplitudes trigger global transport via resonance overlap, per the Chirikov criterion.

These mechanisms apply to mode-coupled systems like mechanical resonators, optomechanics, and wave--mode interactions in plasmas or nonlinear optics, where modulating couplings tunes resonances and energy redistribution. Our scaling relations predict instability thresholds and dynamical transitions.

In the high-frequency limit, potential stiffening---akin to vibrational stabilization---suppresses large excursions and chaos, offering control over resonance and transport via nonlinear modulation.

\section{Conclusion}
The present work establishes a unified analytical and numerical framework for understanding how parametric modulation induces, modulates, and suppresses chaos in the Hénon–Heiles system. Beyond theoretical consistency, the results carry clear physical implications. The derived scaling laws quantify thresholds for separatrix splitting, resonance overlap, and chaos suppression, providing a predictive language for the onset of global transport in Hamiltonian systems. The findings apply directly to astrophysical escape under periodic tidal forcing and to the design of parametrically driven micro/nanoresonators, optical cavities, and plasma traps, where controlled modulation of system parameters governs transitions between stable and chaotic dynamics. The discovery of high-frequency stabilization offers a new route to chaos control via fast parametric drives, complementing standard feedback methods. This analysis holds in the weak-modulation regime ($\varepsilon \ll 1$) near low-order resonances. At higher energies or stronger modulation, higher-order nonlinearities and resonance overlap dominate, necessitating fully numerical treatments. Overall, these results show that modulating nonlinear interaction terms offers a fundamentally different, more direct mechanism for controlling resonance structure and chaos compared to traditional external forcing.

\bibliographystyle{apsrev4-2}  
\bibliography{reference}  

\end{document}